\documentclass{svproc}

\usepackage{amsmath}
\usepackage{mathptmx}
\usepackage{subfig}
\usepackage[labelfont=bf]{caption} 
\usepackage{latexsym}
\renewcommand{\figurename}{\textbf{Fig.}}
\renewcommand{\tablename}{\textbf{Fig.}}

\usepackage{float}
\usepackage{hyperref}
\hypersetup{
    colorlinks=true,
    linkcolor=blue,
    filecolor=blue,      
    urlcolor=blue,
    citecolor=blue,
}
\usepackage{url}
\usepackage{graphicx}
\mainmatter    
\begin{document}

\title{Frequency locking, Quasiperiodicity and Chaos due to special relativistic effects}

\author{Derek C. Gomes \and G. Ambika }

\institute{Department of Physics, Indian Institute of Science Education and Research (IISER) Tirupati , Tirupati - 517507 , India}

\maketitle              
\begin{abstract}
We study quasiperiodic and frequency locked states that can occur in a sinusoidally driven linear harmonic oscillator in the special relativistic regime. We show how the shift in natural frequency of the oscillator with increasing relativistic effects leads to frequency locking or quasi periodicity and the chaotic states that arise due to the increasing nonlinearity. We find the same system can have multi-stable states in the presence of small damping. We also report an enhancement of chaos in the relativistic H\'{e}non-Heiles system.

\end{abstract}
\section{Introduction}

\label{intro}

Most of the studies in nonlinear dynamical systems deal with nonrelativistic regime and chaos exists in such systems due to their inherent nonlinearity. The study of chaos in relativistic systems is an interesting area of research both for its own nature as well as for its applications in many experimental contexts where particle oscillations occur at effectively very high velocities \protect{\cite{Application,Application2}}. The addition of relativistic corrections to linear dynamical systems can induce nonlinearity in them and hence has been reported recently to exhibit chaotic dynamics \protect{\cite{Aniso.,special relativity theory,sp_theory_2,CHORD}}. The classical and quantum dynamics of kicked relativistic particle in a box and driven oscillator are also studied recently \protect{\cite{kicked}}. The 2-dimensional relativistic anisotropic, harmonic oscillator is shown to be chaotic \protect{\cite{Aniso.}}. Also the dynamics of a quartic oscillator at relativistic energies, exhibits bifurcations and chaos and is found to have a  transition to periodic regular motion for large forcing \protect{\cite{quartic}}. The nonlinear dynamics of the constant-period oscillator under external periodic forcing, displays nonlinear resonances and chaos when the driving force is sufficiently strong \protect{\cite{Const.Period}}.

While it is thus established that chaos must appear in most integrable classical systems due to special relativistic corrections to the dynamics, the details of the dynamical states and route to chaos in them are still not fully understood. In our study of the one-dimensional forced harmonic oscillator we show how the natural frequency varies with increasing relativistic effects. We then find that the system exhibits frequency locked and quasiperiodic states as the natural frequency changes. We also indicate a novel route to chaos in this system as the effects of relativistic corrections are tuned. When damping is present, we report that the system exhibits multi stable states due to nonlinear effects induced by relativistic corrections. Also, to the best of our knowledge, our study for the first time on the relativistic version of the H\'{e}non-Heiles system, reports an enhancement of chaotic regions in the relativistic regime.

\section{Relativistic Forced Harmonic Oscillator}

The harmonic oscillator is a popular prototype, widely used in all branches of Physics to understand periodic oscillations as well as to approximate a variety of vibrations in real systems. One of the reasons for this is the extreme simplicity with which it can be used both in classical and quantum regimes giving analytic solutions. This means harmonic oscillator is integrable in classical mechanics and 
analytically solvable in quantum mechanics and this holds true in many dimensions even with damping and forcing.  However, in the special relativistic regime, even the 1-d harmonic oscillator is not integrable and we see the relativistic harmonic oscillator under forcing and damping can give rise to nonlinear dynamical states due to the nonlinearity introduced by special relativistic effects.

The Hamiltonian for the relativistic forced harmonic oscillator (with rest mass unity) is
\begin{equation}\label{eq:1}
H= \sqrt{p^2c^2+c^4}+ \frac{1}{2}kx^{2}+xFcos\omega t
\end{equation}
which leads to the equations of motion
\begin{equation}\label{eq:2}
\frac{dx}{dt}=\frac{p}{\sqrt{1+\frac{p^{2}}{c^{2}}}}
\end{equation}
\vspace{-0.4 cm}
\begin{equation}\label{eq:3}
\hspace{0.6cm} \frac{dp}{dt}=-kx-Fcos\omega t
\end{equation}

It is clear that the above system (all variables in suitable units) is nonlinear , with the spring constant ($k$) , the forcing amplitude ($F$) and the driving frequency ($\omega$). 
As reported by Kim and Lee \protect{\cite{CHORD}}, we control the relativistic effects in the system by treating the speed of light, $c$, as an additional parameter. As we will see in the next section, the relativistic corrections occur as functions of $p/c$ (where $p$ is the momentum) and decreasing the value of $c$ for similar values of $p$ effectively increases the impact of relativistic effects in the system.

\section{Shift in Natural Frequency }

One of the main changes in the system in Eq. (\protect{\ref{eq:1}}) from its non-relativistic counterpart is that its natural frequency is no longer constant with amplitude, just like a nonlinear oscillator. In order to get a better understanding of how this change occurs and the factors that influence this, we expand Eq. (\protect{\ref{eq:2}}) upto first order, to get
\begin{equation}\label{eq:4}
\frac{dx}{dt}=p(1- p^{2}/2c^{2})
\end{equation}
neglecting higher order terms since $c>>p$. In this limit the relativistic effects act like a perturbation as in Eq. (\protect{\ref{eq:4}}). In this context, perturbation approaches can be used as reported in the case of the constant period oscillator \protect{\cite{Const.Period}} using canonical perturbation theory. We use a different(but equivalent) approach developed by Lindstedt \protect{\cite{Lieberman}} to calculate the effect of the perturbation on the natural frequency. Then the expression for the frequency, $\omega_{0,rel}$ (upto first order) is :
\begin{equation}\label{eq:5}
\begin{split}
\omega_{0,rel}=\omega_{0}[1-\frac{3}{16c^{2}}\{p(0)^{2}+{\omega_{0}}^{2}(x(0) +\frac{F}{\omega_{0}^{2}-\omega^{2}})^{2}\}
-\frac{3\omega^{2}F^{2}}{8c^{2}({\omega_{0}^{2}}-\omega^{2})^{2}}]
\end{split}
\end{equation}
Here ($\omega_{0}$) is the natural frequency in the non-relativistic limit ($c\to{\infty}$).
We find that the resultant frequency decreases due to the perturbation. Although we consider the case of a weak relativistic perturbation for the analytical calculation in Eq.(\protect{\ref{eq:5}}) , the approach is suggestive of the parameters that play a role in the shift in frequency. We present this effect more explicitly by computing the power spectra from the numerically obtained time series of the position variable of the system. 

\begin{figure}[h]
\centering
\begin{tabular}{cc}

\textbf{(a)} & \textbf{(b)}\\
\includegraphics[width=0.45\linewidth]{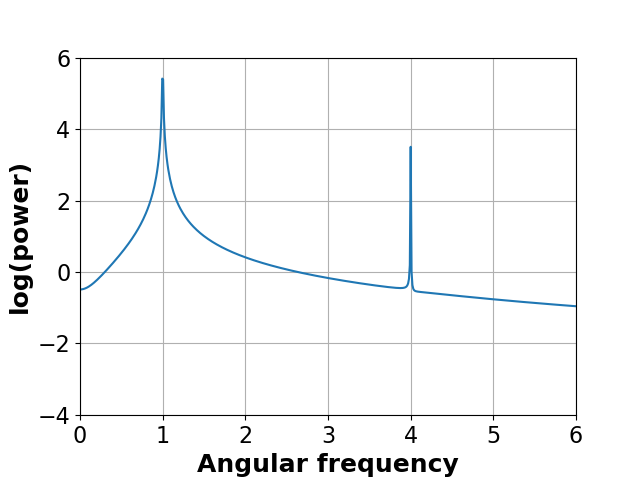} &
\includegraphics[width=0.45\linewidth]{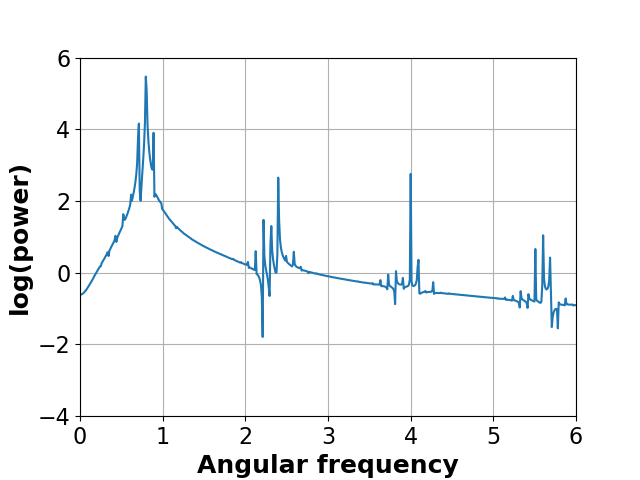} \\
\textbf{(c)}&\textbf{(d)}\\
\includegraphics[width=0.45\linewidth]{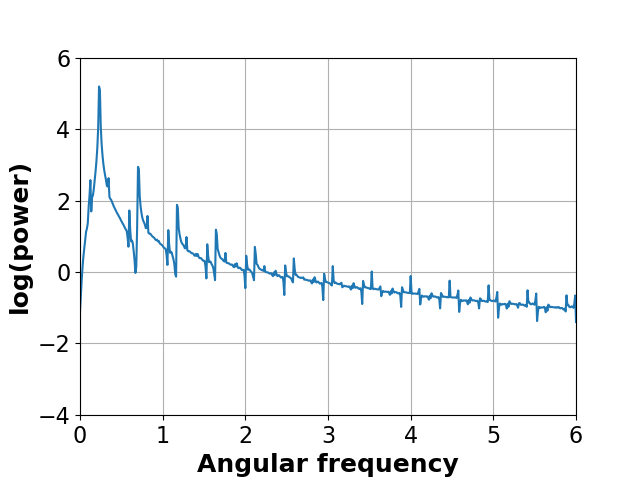} &
\includegraphics[width=0.45\linewidth]{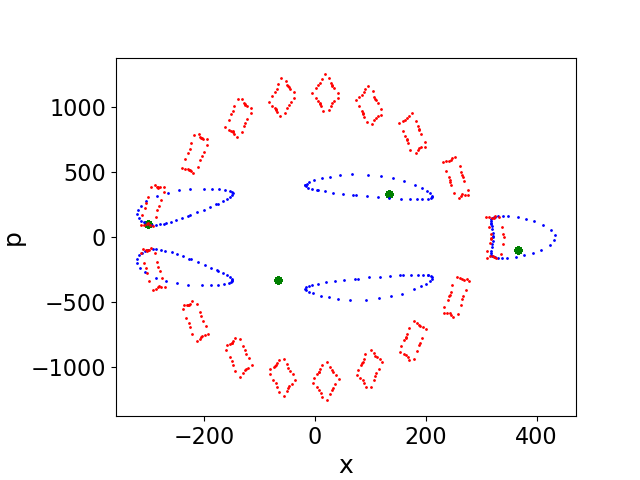}\\
\end{tabular}
\caption{\footnotesize{Power spectra of x(t) for varying $c$. \textbf{a.} $c=300000$ \textbf{b.} $c=300$ \textbf{c.} $c=50$, the other parameters are fixed as $k=1$, $F=500$,$\omega=4$, $x(0)=-300$,$p(0)=100$. \textbf{d.} Poincar\'{e} plot of the phase space showing resonances corresponding to the parameters in the power spectra, colour coded as the red points correspond to  (\textbf{a.}), the blue points to (\textbf{b.}) and the green points (enlarged for visualization) to (\textbf{c.})}}\label{fig:f1}
\end{figure}

In Fig. \ref{fig:f1} we show the power spectra for increasing relativistic effects in the system. It is clear that the natural frequency decreases, while, as expected, the driving frequency (taken as $\omega=4$) remains the same. We note that the greater the relativistic correction (i.e, smaller the parameter $c$), the smaller is the natural frequency corresponding to the largest peak. The dynamical states corresponding to the three values of $c$ used, are clear from the Poincar\'{e} plots in Fig \ref{fig:f1}d .

\begin{figure}[h]
\centering
\includegraphics[width=0.7\linewidth]{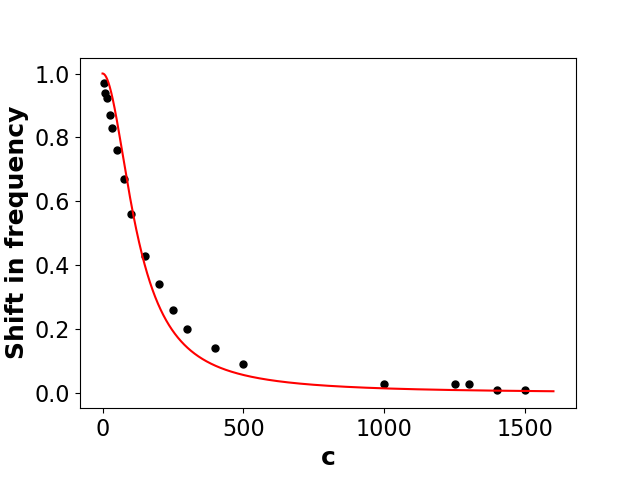}
\caption{\footnotesize{ Shift in the natural frequency of the relativistic harmonic oscillator as a function of $c$ }}\label{fig:f2}
\end{figure}

By tuning $c$ as a parameter, we compute the shift in natural frequency from the power spectra and plot this shift with $c$ in Fig. \ref{fig:f2}. The solid curve in the figure shows the numerical fit to the variation of shift in frequency($\omega_{0}-\omega_{0,rel}$)  $\approx$ $\omega_{0}(1+(c/a)^2)^{-1}$, where $\omega_{0} = 1$. The fitting parameter, $a=122.6$ is estimated by the nonlinear least squares method . We note up to first order, this relation agrees with the nature of variation obtained in Eq.(\protect{\ref{eq:5}}) from the perturbation theory approach.
In the next section, we present how the shift in the natural frequency can lead to different dynamical states in the system.

\section{Frequency Locked and Quasiperiodic states }

\begin{figure}[h]
\centering
\includegraphics[width=0.69\linewidth]{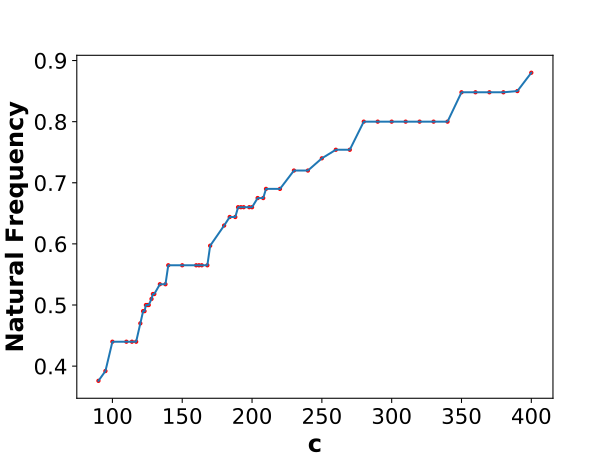}
\caption{\footnotesize{Devil's staircase plot of natural frequency as $c$ is varied ,keeping forcing amplitude constant .The other parameters are the same as in Fig.\ref{fig:f1}}}\label{fig:f3}
\end{figure}
The fact that the natural frequency continuously decreases makes it possible for the system to get into rational and irrational ratios with the driving frequency and therefore the dynamics also leads to frequency locked and quasiperiodic states. The frequency locked states corresponding to rational ratios and the values of $c$ at which they occur are clear from the plateaus in the Devil's Staircase plot in Fig. \ref{fig:f3}. The most prominent plateau is for $w_{0,rel}=0.8$, with the driving frequency at 4 giving a ratio of 5. As the natural frequency continuously decreases, the ratios 6 , 7 , 8  etc. also appear. Also it is equally likely that as the ratios of frequencies vary due to changes in $c$, they enter into irrational ratios and hence can give rise to quasiperiodic dynamics in the system. 

\begin{figure}[h]
\centering
\begin{tabular}{cc}
\textbf{(a)} & \textbf{(b)} \\
\includegraphics[width=0.454\linewidth]{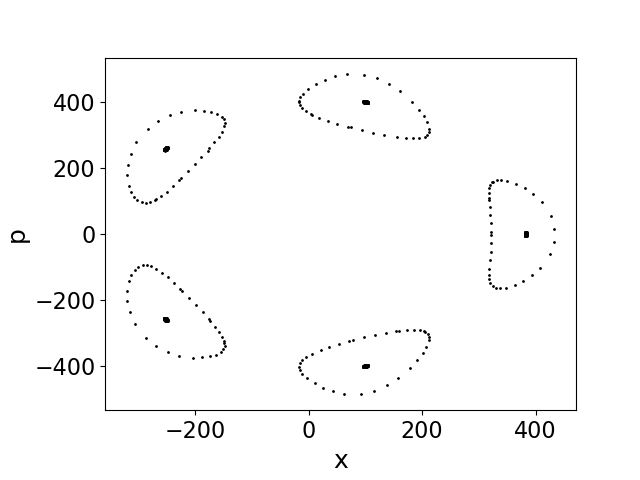} &
\includegraphics[width=0.454\linewidth]{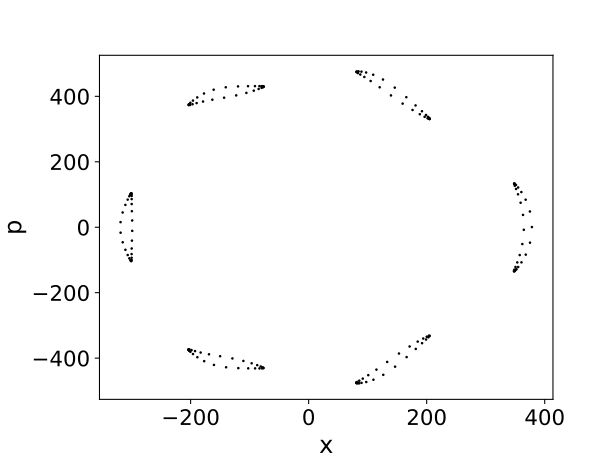} \vspace{0.5cm}\\
\end{tabular}
 \caption{\footnotesize{Poincar\'{e} plots showing frequency locked states with their islands \textbf{a.}  $c=300$, number of resonances, N = 5  \textbf{b.}  $c=200$ , N = 6 }}\label{fig:f4}
\end{figure}

The structure of the trajectories corresponding to both types of dynamical states discussed above are clear from the Poincar\'{e} plots that show trajectories in phase space sampled at the driving frequency. 
The resonances corresponding to the rational ratios of frequencies are shown in Fig. \ref{fig:f4}. As the natural frequency decreases with decrease in $c$, the ratio of frequencies and hence the number of resonances also increase.  
\begin{figure}[h]
\centering
\begin{tabular}{cc}
\textbf{(a)} & \textbf{(b)} \\
\includegraphics[width=0.454\linewidth]{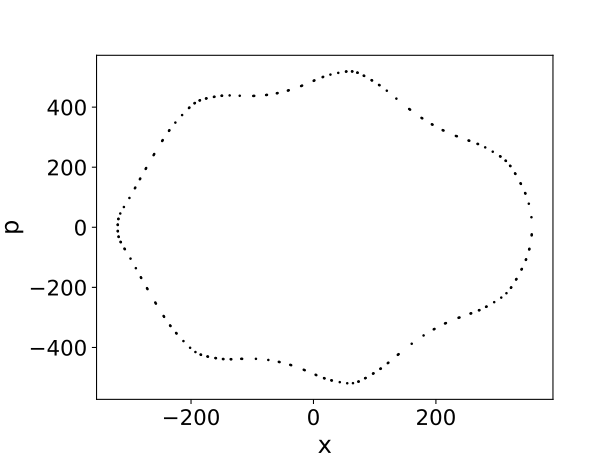} &
\includegraphics[width=0.454\linewidth]{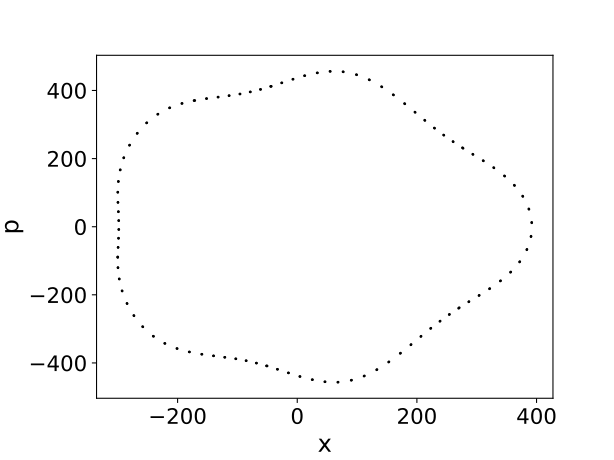} \\
\end{tabular}
 \caption{\footnotesize{Poincar\'{e} plots showing quasiperiodic states. In \textbf{a.} $c=175$ and in \textbf{b.} $c=250$ }}\label{fig:f5}
\end{figure}
We note that the odd numbered (corresponding to odd ratio) resonances are more prominent than their even-numbered counterparts. The occurrence of only odd-numbered resonances (up to first order) has been attributed to the symmetry of the potential in the earlier work on the relativistic driven harmonic oscillator \protect{\cite{CHORD}} as well as in similar relativistic systems \protect{\cite{Const.Period}}. We observe the even-numbered resonances arise from higher-order effects but are confined to very small regions of phase space.
The continuous trajectories shown in Fig.\ref{fig:f5} indicate occurrence of quasiperiodic states for two different values of $c$.

\section{Chaos induced by relativistic effects}

\begin{figure}[H]
\begin{tabular}{cc}

\textbf{(a)} & \textbf{(b)} \\
\includegraphics[width=0.45\linewidth]{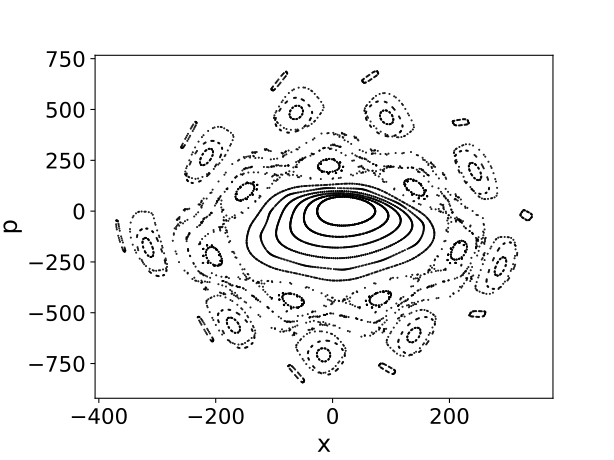} &
\includegraphics[width=0.45\linewidth]{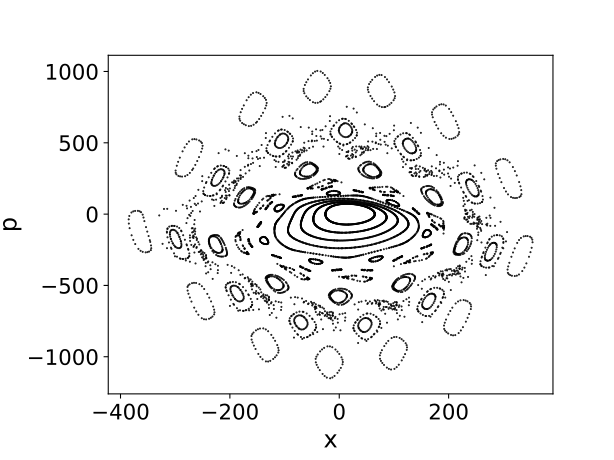} \vspace{0.5cm} \\

\textbf{(c)}& \textbf{(d)} \\
\includegraphics[width=0.45\linewidth]{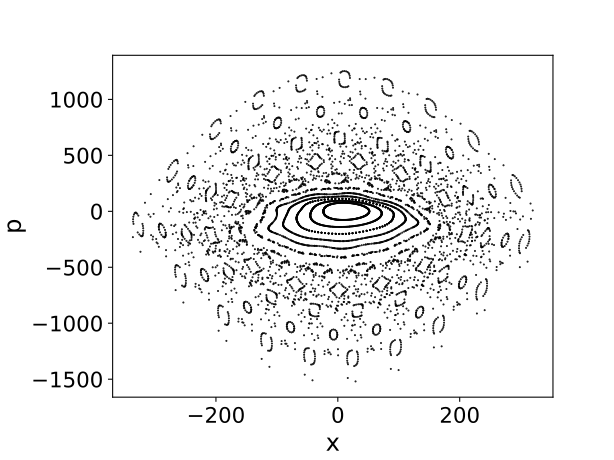} &
\includegraphics[width=0.45\linewidth]{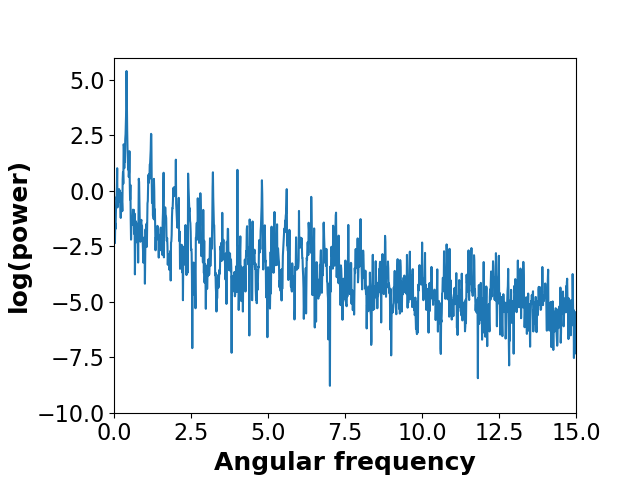}\\
\end{tabular}
     \caption{\footnotesize{Poincar\'{e} plots of the phase space trajectories indicating the appearance of chaotic regions as $c$ is decreased \textbf{a.} $c=100$ \textbf{b.} $c=75$ \textbf{c.} $c=45$, \textbf{d.} Power spectrum for a chaotic trajectory from \textbf{b} }}\label{fig:f6}
\end{figure}

We observe the occurrence of chaos induced by relativistic effects in regions confined between the other types of trajectories . We plot in Fig. \ref{fig:f6} the Poincar\'{e} plots of the phase space of the system with the same set of initial conditions and parameters but reducing the value of $c$. It is clear from Fig \ref{fig:f6} that as $c$ decreases the chaotic regions appear in phase space due to disappearance of the even-resonance states. Thus in Fig. \ref{fig:f6}b. the chaotic trajectory occurs near the N=10 resonance while the nearby trajectories of N=9 and N=11 resonances are still preserved. In Fig \ref{fig:f6}c. where $c$ is decreased even further, only odd resonances are visible in between chaotic regions. To further visualize chaos in this system we plot in Fig. \ref{fig:f6}d. the power spectrum corresponding to a chaotic trajectory in Fig. \ref{fig:f6}b.

\section{Multi-stability in Damped Relativistic Oscillator}

\begin{figure}[h]
\begin{tabular}{cc}
\centering
\textbf{(a)} & \textbf{(b)} \\
\includegraphics[width=0.45\linewidth]{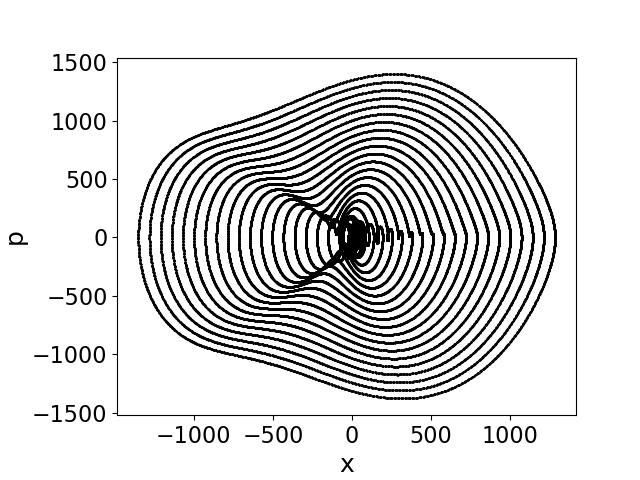} &
\includegraphics[width=0.45\linewidth]{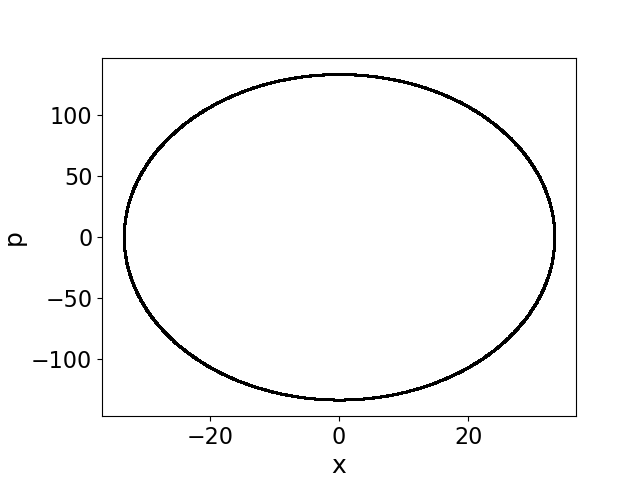} \\
 \textbf{(c)}&\textbf{(d)} \\
\includegraphics[width=0.45\linewidth]{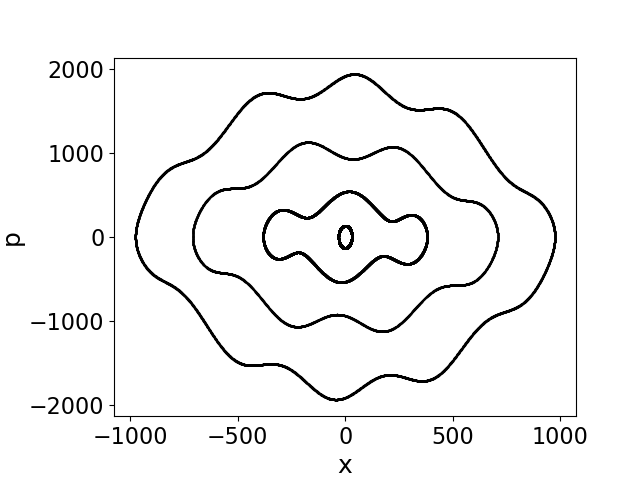} &
\includegraphics[width=0.45\linewidth]{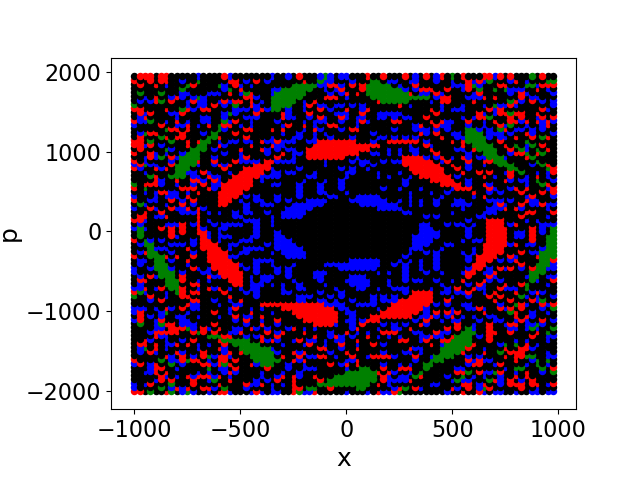}\\	
\end{tabular}
   \caption{\footnotesize{Phase space trajectories starting from the same set of 20 initial conditions for \textbf{a.} undamped nonrelativistic \textbf{b.}, damped nonrelativistic \textbf{c.} damped relativistic oscillator. Here $b=0.01$ and $c=300$. In \textbf{d.} we plot the basin structure for the system where the four different colors correspond to the basins of the four different attractors in the damped relativistic case in \textbf{c.} (the colors black, blue, red and green represent the smallest to largest attractor respectively) }}\label{fig:f7}
\end{figure}

In this section we consider the damped harmonic oscillator in the relativistic regime with the following equations
\begin{equation}\label{eq:6}
\frac{dx}{dt}=\frac{p}{\sqrt{1+\frac{p^{2}}{c^{2}}}}
 ;\frac{dp}{dt}=-kx-bp-Fcoswt
\end{equation}
where the additional term is $-bp$ , with $b$ as the damping parameter. We find that relatively small damping in the system, with relativistic effects leads to multi-stable states, very different from the damped driven nonrelativistic oscillator. This is seen in Fig. \ref{fig:f7} where the phase space structure for the same set of 20 different initial conditions are shown for the undamped nonrelativistic, damped nonrelativistic and damped relativistic cases. In the damped relativistic case, we find the system settles to four different states or attractors. To confirm multi-stability, we present the basin structure corresponding to these four attractors.

\section{Relativistic H\'{e}non-Heiles system: }

\begin{figure}[h]
\begin{tabular}{cc}
\centering
\textbf{(a)} & \textbf{(b)} \\
\includegraphics[width=0.4\linewidth]{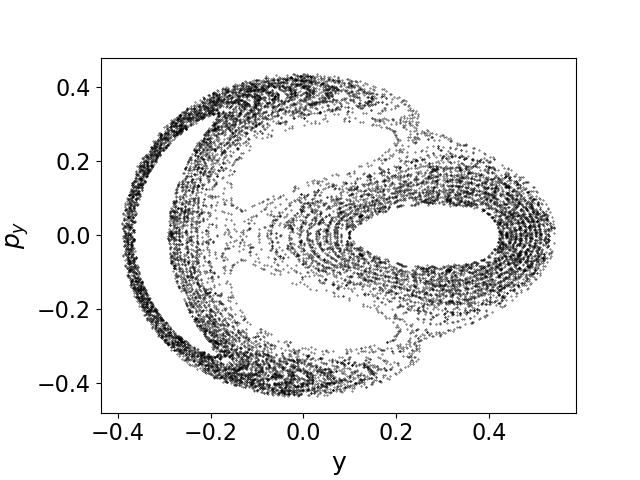} &
\includegraphics[width=0.4\linewidth]{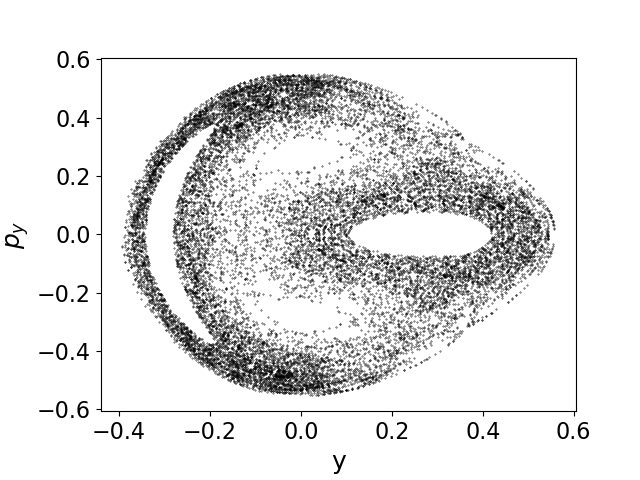} \vspace{0.0cm} \\
\textbf{(c)} & \textbf{(d)}\\
\includegraphics[width=0.4\linewidth]{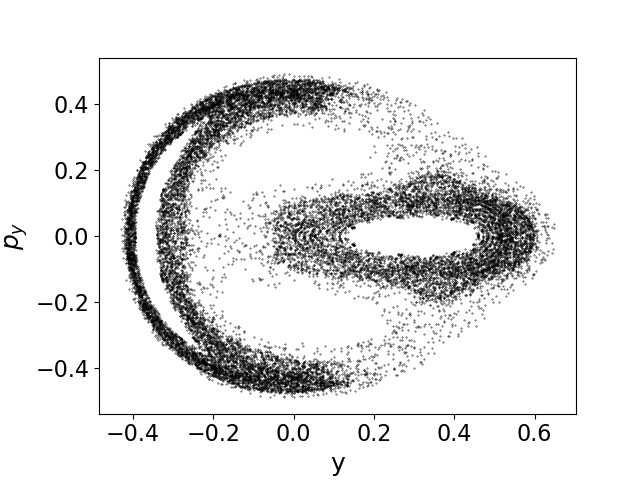} &
\includegraphics[width=0.4\linewidth]{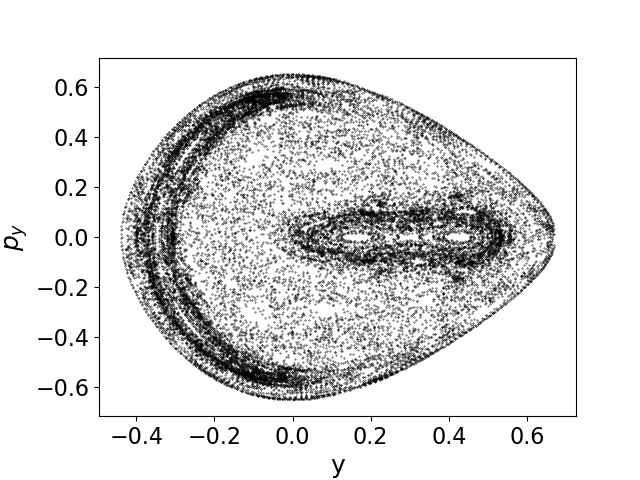}\vspace{0.0cm} \\
\textbf{(e)} & \textbf{(f)}\\
\includegraphics[width=0.4\linewidth]{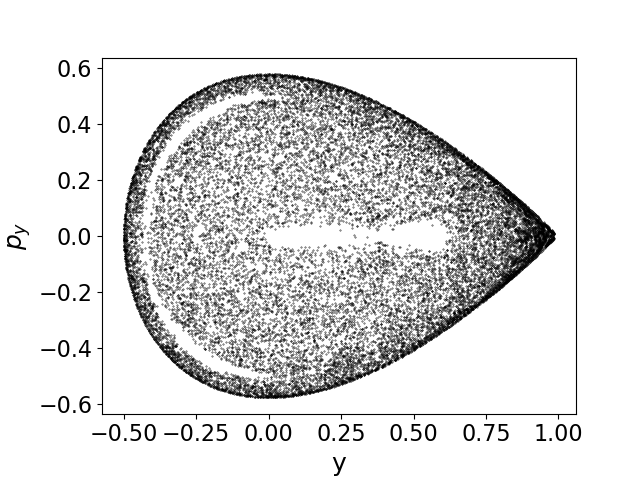} &
\includegraphics[width=0.4\linewidth]{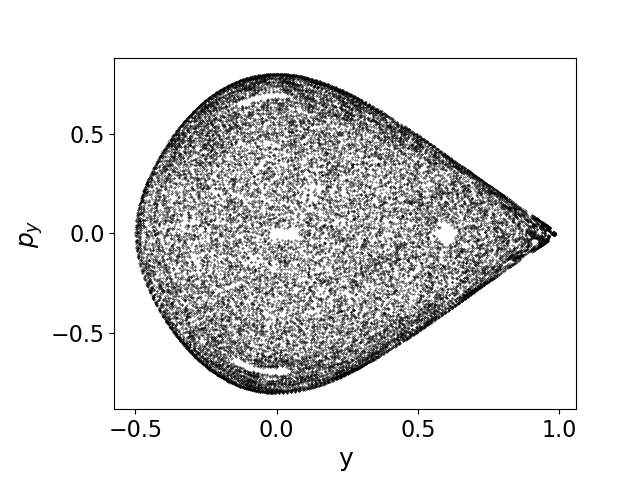} \\
\end{tabular}
     \caption{\footnotesize{Phase space structure in the $y-p_y$ plane for the relativistic H\'{e}non-Heiles system (right) compared with that of the non-relativistic system(left) \textbf{a.} $E=0.10$, non-relativistic \textbf{b.} $E=0.10$, relativistic \textbf{c.} $E=0.125$,non-relativistic \textbf{d.} $E=0.125$, relativistic \textbf{e.} $E=1/6$,non-relativistic \textbf{f.} $E=1/6$, relativistic, where relativistic cases correspond to $c=0.3$ }}\label{fig:f8}
\end{figure}

As another system of interest, we study the effects of special relativity in an intrinsically nonlinear and nonintegrable system, the H\'{e}non-Heiles system. This system models the stellar motion about a galactic center and is known to have a rich dynamics \protect{\cite{Heiles}}. The potential of the system models the galactic potential centred around the galactic center in the $x-y$ plane given by
\begin{equation}\label{eq:7}
V(x,y) = \frac{1}{2}(x^{2}+y^{2}) + \lambda(x^{2}y-\frac{y^{3}}{3})
\end{equation}
The relativistic Hamiltonian is 
\begin{equation}\label{eq:8}
    H= \sqrt{p^2c^2+c^4}-c^{2}+ V(x,y)
\end{equation}
where rest mass is unity and we have subtracted the rest energy($c^{2}$) from the Hamiltonian as it is usually not considered in measuring energy in the non-relativistic case. In general, this has no significance in the equations of motion and as the total energy $E$ is treated as a parameter in studying the H\'{e}non-Heiles system \protect{\cite{Heiles}} and for making comparison with well-studied non-relativistic case, it is important to subtract this contribution. We note that in the limit as $c\to \infty$ the Hamiltonian in Eq. (\protect{\ref{eq:7}}) reduces to the nonrelativistic one. Our main result in this study is that adding relativistic corrections leads to enhanced chaos in the system. This is clear from the phase space structures shown in the $y-p_y$ plane of phase space in Fig. \ref{fig:f8} for both non-relativistic and the corresponding relativistic regimes for three different values of $E$ (we take $\lambda=1$ and $p_y$ corresponds to momentum conjugate to $y$).

\section{Conclusion}

We report the different dynamical states induced by special relativistic effects in a harmonic oscillator with sinusoidal forcing. We present perturbation methods and numerical computations to show how the relativistic corrections shift the natural frequency and consequently generate frequency locked and quasiperiodic states. We also show how reducing the value of $c$ can lead to chaos with the disappearance of even resonances.
In the presence of small damping, the forced harmonic oscillator exhibits multi-stable states with an interesting basin structure. In an inherent nonlinear system like H\'{e}non-Heiles system, we report the enhancement of chaos as the relativistic effects are tuned.

\end{document}